\documentclass[sigconf]{acmart}

\usepackage{multirow}
\usepackage[title]{appendix}
\usepackage{xcolor}
\usepackage{textcomp}
\usepackage{algorithm}
\usepackage{algorithmicx}
\usepackage{algpseudocode}
\usepackage{listings}
\usepackage{pgfplots}
\usepackage{fancybox}
\usepackage{xcolor}
\usepackage{enumitem}

\newcommand{\update}[1]{#1}

\copyrightyear{2026}
\acmYear{2026}
\setcopyright{cc}
\setcctype{by}
\acmConference[AI-SQE '26]{1st International Workshop on AI for Software Quality Evaluation - Judgment, Metrics, Benchmarks, and Beyond}{April 12--18, 2026}{Rio de Janeiro, Brazil}
\acmBooktitle{1st International Workshop on AI for Software Quality Evaluation - Judgment, Metrics, Benchmarks, and Beyond (AI-SQE '26), April 12--18, 2026, Rio de Janeiro, Brazil}
\acmPrice{}
\acmDOI{10.1145/3786175.3788343}
\acmISBN{979-8-4007-2407-7/2026/04}

\begin{document}

\title[Precision or Peril: A PoC of Python Code Quality]{Precision or Peril: A PoC of Python Code Quality from Quantized Large Language Models}

\author{Eric L. Melin}
\email{ericmelin@u.boisestate.edu}
\affiliation{
  \institution{Boise State University}
  \city{Boise}
  \state{Idaho}
  \country{USA}
}
\authornote{These authors contributed equally to this work.}

\author{Adam J. Torek}
\email{adamtorek@u.boisestate.edu}
\affiliation{
  \institution{Boise State University}
  \city{Boise}
  \state{Idaho}
  \country{USA}
}
\authornotemark[1]

\author{Nasir U. Eisty}
\email{neisty@utk.edu}
\affiliation{
  \institution{University of Tennessee}
  \city{Knoxville}
  \state{Tennessee}
  \country{USA}
}

\author{Casey Kennington}
\email{caseykennington@boisestate.edu}
\affiliation{
  \institution{Boise State University}
  \city{Boise}
  \state{Idaho}
  \country{USA}
}

\renewcommand{\shortauthors}{Melin, et al.}


\begin{abstract}
\textbf{\textit{Context:}} Large Language Models (LLMs) like GPT-5 and LLaMA-405b exhibit advanced code generation abilities, but their deployment demands substantial computation resources and energy. Quantization can reduce memory footprint and hardware requirements, yet may degrade code quality.  
\textbf{\textit{Objective:}} This study investigates code generation performance of smaller LLMs, examines the effect of quantization, and identifies common code quality issues as a proof of concepts (PoC).  
\textbf{\textit{Method:}} Four open-source LLMs are evaluated on Python benchmarks using code similarity metrics, with an analysis on 8-bit and 4-bit quantization, alongside static code quality assessment.  
\textbf{\textit{Results:}} While smaller LLMs can generate functional code, benchmark performance is limited. Quantization impacts are variable, and generated code exhibits quality and maintainability concerns.  
\textbf{\textit{Conclusions:}} LLM-generated code should be carefully validated before integration into software projects.
\end{abstract}


\maketitle

\section{Introduction}
Advancements in artificial intelligence and machine learning have reshaped software engineering with large language models (LLMs) demonstrating strong potential to automate tasks such as code completion, analysis, test case generation, and documentation~\cite{LLM_SoftwareEngineeringSurvey_2023}.
Recent models like GPT-5, LLaMA-405B, Qwen 3 Coder, and LLaMA 4 Maverick show impressive capabilities, but also raise concerns about code quality and deployment.
LLM generated code may be incomplete, inefficient, or contain runtime, logical, or security errors~\cite{siddiq2023generate}. 
Their billions of parameters also demand significant memory and computation, limiting feasibility for smaller organizations.

To address these challenges, both industry and academia employ evaluation metrics, benchmarks, and model compression techniques. 
Benchmarks such as HumanEval~\cite{liu2024your}, \update{Mostly Basic Python Problems} (MBPP) ~\cite{MBPP_2021}, and SWE-Bench~\cite{SWE_Bench_2024}, along with similarity metrics like CodeBLEU~\cite{ren2020codebleu}, provide quantitative assements of code generation quality.
Static and dynamic analysis tools, such as SonarQube~\cite{sonarqube-doc}, unit tests, integration tests, and vulnerability scanners, detect code smells and other quality issues, enabling a comprehensive understanding of generated code. 

Quantization is a widely used technique to reduce memory requirements~\cite{AWQ_2023, GPTQ_2023}, but it may degrade output quality, a phenomenon is known as ``quality loss''.
Despite widespread adoption of evaluation tools and quantization, the interaction between model compression and code quality, as well as the relationship  between different evaluation metrics, remains underexplored.

This study addresses these gaps by systematically evaluating four open-source LLMs across three dimensions: benchmark performance, similarity metrics, and static analysis. 
We test both unquantized and quantized versions of each selected model (4-bit and 8-bit), using HumanEvalPlus and MBPP Plus as functional benchmarks, CodeBLEU for similarity scoring, and SonarQube for maintainability analysis. 
Although the specific models studied represent an earlier generation of LLMs, the challenges we observe as a proof of concept, poor functional reliability, mixed quantization effects, and widespread maintainability concerns, remain highly relevant for current and future systems. 
Our findings highlight risks that persist across model families and offer practical lessons for developers and researchers adopting LLMs in production.

We pose the following research questions to drive our study:
\begin{itemize}[leftmargin=1em]
    \item \textbf{RQ1:} How well does each LLM perform on HumanEvalPlus and MBPP Plus, and how do they compare with the CodeBLEU scores of each model?
    \item \textbf{RQ2:} What is the effect of quantization on LLM performance on benchmark scores and CodeBLEU scores?
    \item \textbf{RQ3:} What kinds of code quality issues (\update{i.e.}, code smells, readability issues, maintainability issues) does unquantized \update{and quantized} LLM-generated code have, and what can automatic and manual analysis show about these issues?
\end{itemize}

The key contributions of this paper are:
\begin{itemize}[leftmargin=1em]
    \item Analysis of LLM performance on benchmarks, code similarity scores, and static analysis metrics.
    \item Evaluation of the effects of quantization on functional correctness and maintainability.
    \item Identification of common code quality issues in LLM-generated Python code.
    \item A public GitHub repository with all data and code\footnote{https://figshare.com/s/5e6ecebe07829ee8b106}.
\end{itemize}

\section{Related Work}
Recent studies have examined how to apply LLMs to software engineering tasks. Much of this research is focused on benchmarks. Some studies have proposed new metrics~\cite{Du_Luu_Ji_Ng_2024} and datasets~\cite{Zhong_Wang_2024} for evaluating the quality and functionality of LLM-generated code. Other works, such as SALLMS~\cite{Siddiq_Santos_2023} and CyberSecEval~\cite{Bhatt_Chennabasappa_2023}, have introduced comprehensive benchmarks for systemically assessing LLM outputs. Additional work has examined LLM-generated software quality from a user perspective by integrating LLMs into software engineering workflows~\cite{Vaithilingam_2022, Nguyen_Babe_Zi_Guha_Anderson_Feldman_2024}.

While prior work has made progress on individual dimensions: security, efficiency, robustness, or user experience, few studies have examined how different evaluation metrics relate to one another or how quantization on smaller LLMs may impact software quality.
Our work addresses this gap by combining functional benchmarks, similarity scores, and static analysis, and by explicitly comparing unquantized and quantized models.
\section{Methodology}

\subsection{Methodology Overview}

Fig.~\ref{fig:method_diagram} shows an overview of our methodology and experimental pipeline used to answer our proposed RQs. Below is a brief description of each step:
\begin{figure}[ht]
    \centering
    \includegraphics[width=0.9\columnwidth]{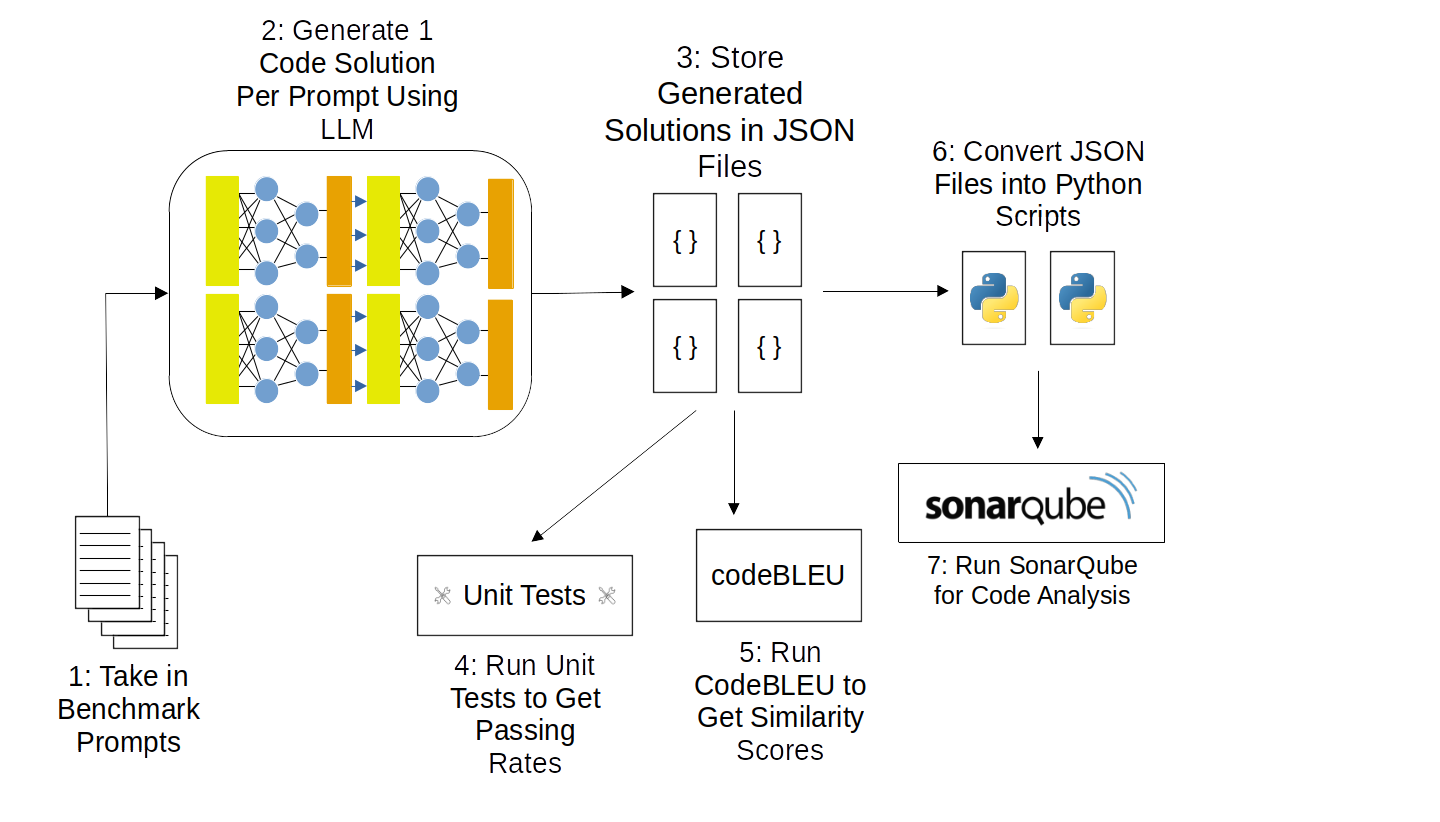}
    \caption{\update{Methodology for Code Quality Evaluation.}}
    \label{fig:method_diagram}
\end{figure}

\begin{itemize}[leftmargin=1em]
    \item \textbf{Step 1 - Take in Benchmark Prompts:} This first step extracts each prompt from both the HumanEvalPlus and MBPP Plus benchmarks and prepares them to be given to the quantized/unquantized version of each LLM. We did not do any prompt engineering before sending the prompt into the LLM.
    \item \textbf{Step 2 - Generate 1 Code Solution Per Prompt Using the LLM:} Here, we generate a code solution for each prompt using the LLM itself. To do this, we first tokenize the prompt using the LLM's tokenizer, feed the prompt to the LLM, and decode the output once its generation is complete.
    \item \textbf{Step 3 - Store Generated Solutions:} This step extracts each generated solution and saves them to JSON files. We saved all the results for a given benchmark for further analysis in steps 4 - 7.
    \item \textbf{Step 4 - Run Unit Tests to Get Passing Rates:} This step is the first in our analysis phase. Here, we run the unit tests provided with each benchmark on the generated solutions. No preprocessing was done in this step.
    \item \textbf{Step 5 - Run CodeBLEU to get Similarity Scores:} In this step, we ran CodeBLEU to get the similarity scores between the generated solutions and the human-written solutions provided in both benchmarks. No preprocessing was required.
    \item \textbf{Step 6 - Convert solutions to Python Files:} Before we run our SonarQube analysis, we convert the solutions into Python files. To do this, we extracted each solution from the JSON files and write them into a separate Python file. This results in a single Python file for each solution that we could then use for SonarQube analysis in step 7.
    \item \textbf{Step 7 - Run SonarQube for Code Analysis:} Once the generated code is written to Python files, we run SonarQube on all the Python files for each of the benchmarks.  SonarQube analysis results in many different metrics, including code smells, vulnerability issues, maintainability and reliability problems, code efficiency issues, and the estimated time to fix all of these issues. 
\end{itemize}

\update{We followed the same 7-step procedure shown in the diagram for 4-bit quantized, 8-bit quantized, and unquantized base LLMs.}

\subsection{Chosen LLMs}

We selected four open source LLMs to use for our testing. Due to limited memory constraints, our LLMs had to be able to fit on a single GPU with and without quantization. \update{Details about the GPU we used, as well as our experimental environment, can be found in Section \ref{sec:experimental_setup}.} Our quantization analysis also removed proprietary LLMs from our study, as the quantization state of those LLMs could not be controlled through user settings. These constraints limited us to using open source LLMs with less than 10 billion parameters. With these constraints, we selected four different open-source LLMs that were trained and fine-tuned on code-specific tasks using different regimes, architectures, and datasets (explained below for each model). These are the four LLMs we analyzed in this study:

\begin{enumerate}[leftmargin=1em]
    \item \textbf{WizardCoder 7B:} A 7B code LLM fine-tuned from Llama using Evol-Instruct~\cite{Luo_Xu_Zhao_Sun_Geng_Hu_Tao_Ma_Lin_Jiang_2023}, which iteratively increases prompt difficulty through adversarial training. We used the 7B version due to memory constraints.  
    \item \textbf{Mistral Instruct 7B:} A 7B LLM pretrained on public datasets and fine-tuned for prompting~\cite{Mistral_2023}. Optimized for efficiency and capable of strong code generation despite not being code-specific. Chosen to compare against code-specialized LLMs.
    \item \textbf{StarCoder 2 7B:} A 7B code LLM trained on The Stack V2, a trillion-token dataset of GitHub artifacts~\cite{StarCoder2_2024}. Serves as a strong base for coding tasks; we used the 7B version as the largest GPU-compatible model.
    \item \textbf{LlamaCode 7B:} A 7B code LLM fine-tuned from Llama-2 on 500B tokens of code~\cite{CodeLlama_2024}, with extended context for improved code generation. Selected for GPU compatibility.
\end{enumerate}

\subsection{LLM Code Generation}
For LLM code generation, we use two benchmarks as sources for prompts: HumanEvalPlus \cite{chen2021evaluating} and Mostly Basic Python Problems (MBPP) Plus \cite{MBPP_2021}. Each benchmark contains a set of Python programming problems in the form of prompts that an LLM has to solve. Once the LLM completes all the prompts, they are evaluated through a series of unit tests, and the average passing rate across all the test cases is returned as the final score for both benchmarks. A human-written solution is included for all test cases as a baseline for comparison. HumanEvalPlus has 169 medium-difficulty Python problems with 40-80 test cases per problem. MBPP Plus has roughly 800 easy Python problems that have about 10-30 test cases per problem. The test suites of both HumanEvalPlus and MBPP Plus were expanded from HumanEval and MBPP, respectively, through a combination of context-free grammars (CFGs) and LLMs. We used the expanded test cases for our benchmark results. An example of the HumanEvalPlus prompt can be seen in Listing~\ref{lst:Listing1} and a corresponding generated LLM solution in Listing \ref{lst:Listing2}.

\lstset{
    basicstyle=\ttfamily\footnotesize, 
    backgroundcolor=\color{lightgray!20}, 
    keywordstyle=\color{blue}\bfseries, 
    commentstyle=\color{green!60!black}, 
    stringstyle=\color{red}, 
    numbers=left, 
    numberstyle=\tiny\color{gray},
    frame=single, 
    breaklines=true, 
    captionpos=b, 
}

\begin{lstlisting}[language=Python, caption={HumanEvalPlus Prompt \#2}, label={lst:Listing1}]
def truncate_number(number: float) -> float: 
""" Given a positive floating point number, it can be decomposed into and integer part (largest integer smaller than given number) and decimals (leftover part always smaller than 1). 

Return the decimal part of the number. 
>>> truncate_number(3.5) 0.5 """
\end{lstlisting}

\begin{lstlisting}[language=Python, caption={Mistral7B Generated Solution}, label={lst:Listing2}]
def truncate_number(number: float) -> float: 
""" Given a positive floating point number, it can be decomposed into and integer part (largest integer smaller than given number) and decimals (leftover part always smaller than 1). 

Return the decimal part of the number. 
>>> truncate_number(3.5) 0.5 """
    return number % 1
\end{lstlisting}

For quantization testing, we used Activation Aware Weight Quantization (AWQ) \cite{AWQ_2023} to quantize our models using 4-bit and 8-bit quantization. This quantization method works by finding the most important parameters in an LLM through activation and quantizing the entire model based on those weights, which results in compressed parameters that preserve much of the LLM's original performance while still reducing its memory requirements significantly. Another benefit of AWQ quantization is that once the activation testing is complete, the LLM's parameters can be quantized into 8-bit and 4-bit spaces without further testing. We chose this as our quantization method due to its simple setup, use, and because it achieves a good balance between increased LLM generation speed, performance, and decreased memory requirements \update{\cite{AWQ_2023}}.

Once the benchmarks and quantization method were selected, we had each LLM generate one solution for each programming problem in the benchmarks. We used the \update{unmodified starter prompt for each problem} and had each LLM perform a code completion task based on that prompt. Once all solutions were generated for both HumanEvalPlus and MBPP Plus, results were saved. We repeated this process for each LLM with 4-bit and 8-bit quantization, as well as no quantization.  This resulted in eight generation runs across all LLM quantization settings, and 24 solution files in total. These solution files became the basis for all of our evaluations and analysis after code generation was complete.

\subsection{Benchmark and CodeBLEU Scoring}
Once the solution files were generated, we then evaluated all the LLMs and their quantization results using the unit tests provided in the HumanEvalPlus and MBPP Plus benchmarks. For both benchmarks, we used the pass@1 score, which runs all the unit tests on a single generated solution and reports the test passing rate per solution. We recorded the test passing rate for the expanded test cases in our evaluation into a single solution file for analysis and comparison to other metrics. Our benchmark results were used as a baseline to compare our other evaluation metrics and were utilized to investigate the impact of AWQ on the LLM's weights. We ran our benchmarks on all four LLMs \update{utilizing two quantization settings, and an unquantized setting}, resulting in 12 benchmark scores for HumanEvalPlus and 12 benchmark scores for MBPP Plus.

We used CodeBLEU as our similarity metric for comparing LLM-generated and human written code. Both HumanEvalPlus and MBPP Plus have correctly implemented human solutions for all the programming problems in both benchmarks. Since CodeBLEU is a dataset-level benchmark that uses data flow and syntax tree evaluations across an entire codebase as part of its comparison process, we ran it on all the generated and human written solutions at once rather than one solution at a time.

\subsection{SonarQube Analysis}
\label{sec:sonarqube analysis}
We used SonarQube as our static analysis tool for our experimentation. SonarQube is a static analysis framework that can run many different kinds of analyses, including code smells, security vulnerabilities, linting, maintainability issues, readability issues, code inefficiencies, and estimated time to fix issues. 
We selected SonarQube as our static analysis tool because of its ease of use, variety of metrics we could use in our analysis, \update{and its proven effectiveness in research.
For example, Mat'ašová et al.~\cite{mat2024enhancing} demonstrated SonarQube’s precise detection of code smells and bugs in student assignments, which enhanced code quality and maintainability. 
Similarly, Gurung et al.~\cite{gurung2024static} showcased its ability to identify energy-related code smells in Java loops, while De Luca et al.~\cite{de2024automatic} highlighted its effectiveness in detecting code smells and architectural anti-patterns in student Java projects, improving maintainability and adherence to design principles. 
These findings underscore SonarQube’s reliability and versatility for analyzing all the LLM-generated Python files.} We used the Community Edition of SonarQube for our analysis since the code base of our generated solutions is not large, and we did not require additional features of the paid editions.

Since the generation of the LLM code had been stored in individual solution files for each unquantized and quantized LLM dataset pair, we had a total of 24 different solution files containing a total of 6,756 JSON objects containing both the dataset question ID and the generated LLM solution. We then needed to parse through each object to create a python file from each in order to run SonarQube analysis on the generated code. To do this, we created and used a Python script that parsed each line and translated the corresponding python code as an individual python file. Once complete, we then fed each batch of Python files from each benchmark into SonarQube for analysis. 

\update{SonarQube ~\cite{sonarqube-doc} defines clean code as code that has the following attributes: consistency, intentionality, adaptability, and responsibility. Consistency refers to the code being written in a uniform and conventional way. All the code should look similar and follow a regular pattern, even with multiple contributors at different times. \textit{Consistent code is formatted, conventional, and identifiable.} Intentionality refers to code that is precise and purposeful. Every instruction makes sense, is adequately formed, and clearly communicates its behavior. \textit{Intentional code is clear, logical, complete, and efficient.} Adaptability means that code is structured to be easy to evolve and develop with confidence. It makes extending or repurposing its parts easy and promotes localized changes without undesirable side-effects. \textit{Adaptable code is focused, distinct, modular, and tested.} Responsibility considers that the code takes into account its ethical obligations on data, as well as societal norms. \textit{Responsible code is lawful, trustworthy, and respectful.}}

\update{SonarQube also analyzes code in terms of software quality, based on SonarQube's taxonomy~\cite{sonarqube-doc}. SonarQube categorizes issues into three primary dimensions: \textit{Security}, \textit{Reliability}, and \textit{Maintainability}. \textit{Security} refers to vulnerability issues that may expose the code to exploitation or data breaches. \textit{Reliability} captures bug defects that can cause incorrect behavior or crashes. \textit{Maintainability} includes code smells patterns that, while not necessarily incorrect, indicate deeper structural or stylistic problems that hinder future development. All the preceeding SonarQube definitions are utilized for analysis in section \ref{sec:RQ3}.}

\subsection{Experimental Setup}
\label{sec:experimental_setup}
We used a single machine for LLM code generation, equipped with an Nvidia RTX 3090 GPU (24BG VRAM), a 16-core / 32-thread AMD Ryzen CPU, and 64 GB RAM. 
The system ran Ubuntu 22.04, Python 3.11, PyTorch 2, and CUDA 12.1.
Model weights and quantized variants were obtained via Hugging Face, using autoawq for quantization. 
Evaluations employed the HumanEvalPlus and CodeBLEU Python libraries and SonarQube Community Edition version 25.5.0. 
All LLMs and quantization settings used a temperature of 0.1 and top-P of 0.9  to ensure experimental consistency.
\section{Results}

This section presents the key findings of our study, addressing the research questions and providing a detailed analysis of the benchmark performance, code similarity, and static analysis. We analyze the effects of quantization on the code quality generated by smaller LLMs, focusing on their utility and limitations in real-world software engineering applications. \update{A summary of the benchmarks and model settings tested in this study can be seen in Table \ref{tab:experiment-setup}}.

\begin{table}[h]
\centering
\scriptsize
\caption{\update{Summary of Benchmarks \& Quantization Levels}}
\label{tab:experiment-setup}
\begin{tabular}{|l|c|p{0.25\columnwidth}|}
\hline
\textbf{Benchmark} & \textbf{Quantization Levels} & \textbf{\# of Test Cases in Benchmark} \\
\hline
HumanEvalPlus        & Unquantized, 8-bit, 4-bit  & 164  \\
MBBP Plus        & Unquantized, 8-bit, 4-bit  & 399  \\
\hline
\end{tabular}
\end{table}

\subsection{\textbf{RQ1: LLM performance}}

\begin{table}[htbp]
    \centering
     \caption{Benchmark Results For Expanded Test Cases}
    \scriptsize
    \begin{tabular}{p{1.0cm}|p{0.5cm}p{0.5cm}p{0.5cm}p{0.5cm}p{0.5cm}p{0.5cm}p{0.5cm}p{0.5cm}}
    \toprule
     Model & \multicolumn{2}{c}{CodeLlama 7B} & \multicolumn{2}{c}{Mistral Instruct 7B } & \multicolumn{2}{c}{WizardCoder 7B} & \multicolumn{2}{c}{Starcoder2 7B} \\
     \toprule
     Benchmark & Human EvalPlus & MBPP Plus & Human EvalPlus & MBPP Plus & Human EvalPlus & MBPP Plus & Human EvalPlus & MBPP Plus \\
\midrule
4 bit & 3.66\% & 12.03\% & 14.02\% & 34.09\% & 21.34\% & 26.07\% & 1.22\% & 1.00\% \\
8 bit & 2.44\% & 11.28\% & 14.63\% & 34.09\% & 19.51\% & 24.81\% & 0.00\% & 0.75\% \\
unquantized & 3.66\% & 11.03\% & 25.61\% & 27.57\% & 27.44\% & 28.32\% & 0.61\% & 1.00\% \\
\bottomrule
\end{tabular}
    \label{tab:evals_plus}
\end{table}

\update{Table \ref{tab:evals_plus} shows our LLM evaluation results in 4-bit quantization, 8-bit quantization, and unquantized formats on HumanEvalPlus and MBPP Plus benchmarks. Scores range from 0\% to 100\%, with higher values indicating better functionality. Overall, all models performed poorly, with the best performing model, Mistral Instruct 7B, scoring at 34.1\% on the MBPP Plus benchmark, while the best scoring model on the HumanEvalPlus benchmark, WizardCoder 7B, scored at a 27.6\% on all test cases. The worst performing model is StarCoder 2 7B, scoring above 1.2\% at best. CodeLlama 7B did better than StarCoder 2 7B, scoring at least 3.66\% on HumanEvalPlus and 11\% on MBPP Plus. WizardCoder 7B did the second best across benchmarks, with Mistral Instruct 7B doing the best overall.}

\update{Table~\ref{tab:codebleu} contains CodeBLEU results from the LLMs and quantization settings we tested for both HumanEvalPlus and MBPP Plus benchmarks. Higher CodeBLEU scores indicate a greater resemblance between LLM-generated output and human-written code. Nearly all of the models tested overall scored between the 0.20 and 0.30 on the CodeBLEU distance metric, showing that all LLMs generated code that at least slightly resembled the human-written solutions for both benchmarks. The highest CodeBLEU scores were achieved by WizardCoder 7B, followed by Mistral Instruct 7B. Once again, StarCoder 2 7B received the lowest CodeBLEU scores, indicating that it deviates from human written code the most, with CodeLlama 7B achieving the second lowest scores out of all the models tested.}

\update{Overall, all of the models performed quite poorly on the pass@1 test scores of both the HumanEvalPlus and MBPP Plus benchmarks, but achieved higher CodeBLEU similarity scores across the board. These poor results indicate that smaller, open source LLMs perform noticeably worse than their much larger counterparts. This is likely due to their limited parameter counts, which may be too small or shallow to fully capture many of the patterns expressed in the datasets they were pretrained and fine-tuned on. These poor results also highlight the need for prompt engineering, self repair, and multi-shot generation. There is some correlation between CodeBLEU scores and our HumanEvalPlus and MBPP Plus test scores, however the results for CodeBLEU on average are much higher than the test case scores for all LLMs tested. This result may be the product of how these LLMs were trained, which likely prioritizes code resemblance and pattern matching over functionality or passing rates \cite{Bender2020-ck}. These results also show that code similarity metrics like CodeBLEU are not accurate or useful substitutes for unit tests in the case of evaluating LLM code quality.}

\colorbox{gray!20}{%
  \begin{minipage}{0.93\columnwidth}
  \textbf{RQ1 Findings}: \update{Our poor LLM-generated results on the HumanEvalPlus and MBPP Plus test suites demonstrates the need for prompt engineering, self-repair techniques, and multiple rounds for LLM code generation. Our higher CodeBLEU similarity scores may indicate that LLMs, without any prompting or guidance, prioritize code similarity and pattern matching over code functionality and edge cases. Additionally, CodeBLEU scores are not a substitute for unit tests or test suites.}
  \end{minipage}
}

\subsection{\textbf{RQ2: Effect of quantization on LLM performance}}

\update{Tables \ref{tab:evals_plus} shows the test suite scores of our AWQ quantized LLMs quantized to 8 bit and 4 bit settings relative to their unquantized versions. Our results show that quantization has mixed effects on LLMs for both the HumanEvalPlus and MBPP Plus benchmarks. Models like StarCoder 2 7B and CodeLlama 7B seem to not be affected much by quantization, but the other two LLMs we tested, Mistral Instruct 7B and WizardCoder 7B, are affected quite strongly. CodeLlama and StarCoder 2 are only affected by about 1\% to 2\% across all benchmarks. Our results for Mistral Instruct and WizardCoder, on the other hand, have differences of 5\% and 6\% for their quantized versions relative to their unquantized ones. The quantized Mistral Instruct, lost a full 13\% on its performance for HumanEvalPlus, scoring a mere 14\% compared to the base LLM, yet increased its performance 7\% to 34.1\% on the MBPP Plus benchmark. WizardCoder was less affected by quantization, with only a 4\% drop in its MBPP scores against its unquantized state in 8 bit quantization and a more severe 8\% drop in performance on the HumanEvalPlus results.}

\begin{table}[htbp]
    \centering
     \caption{CodeBLEU Scores comparing Generated VS Baseline Similarity}
     \scriptsize
    \begin{tabular}{p{1.0cm}|p{0.5cm}p{0.5cm}p{0.5cm}p{0.5cm}p{0.5cm}p{0.5cm}p{0.5cm}p{0.5cm}}
    \toprule
     Model & \multicolumn{2}{c}{CodeLlama 7B} & \multicolumn{2}{c}{Mistral Instruct 7B } & \multicolumn{2}{c}{WizardCoder 7B} & \multicolumn{2}{c}{Starcoder2 7B} \\
     \toprule
     Benchmark & Human EvalPlus & MBPP Plus & Human EvalPlus & MBPP Plus & Human EvalPlus & MBPP Plus & Human EvalPlus & MBPP Plus \\
    \midrule
4 bit & 0.247184 & 0.238342 & 0.244284 & 0.213382 & 0.287238 & 0.262609 & 0.203607 & 0.257308 \\
8 bit & 0.239826 & 0.236393 & 0.243097 & 0.213401 & 0.283712 & 0.260337 & 0.212709 & 0.258715 \\
unquantized & 0.239987 & 0.241254 & 0.262793 & 0.244285 & 0.279789 & 0.259317 & 0.188234 & 0.257682 \\
    \bottomrule
    \end{tabular}
   
    \label{tab:codebleu}
\end{table}

\update{Our CodeBLEU distance metrics as shown in Table \ref{tab:codebleu} are much less effected by quantization for all of our LLMs. All of the LLMs remained consistent between their 8-bit and 4-bit quantized versions relative to their unquantized versions. Most of the differences in these scores can be attributed to random variations in LLM output and small changes caused by AWQ's quantization method. Our results shows that code-based LLMs will generate code with similar structure, regardless of the quantization setting used. Our findings demonstrate that quantization has little effect on structure and syntax of LLM generated code while having strong effects on the functionality and logic of that generated code depending on the model, some positive and some negative. Overall, our results indicate that quantization has little effect on syntax and structure in LLM-generated code while having mixed effects on the functionality and logical correctness of said code.}

\colorbox{gray!20}{%
  \begin{minipage}{0.93\columnwidth}
    \textbf{RQ2 Findings:} \update{ Quantization has mixed effects on the functionality and logic of LLM-generated code while having little effect on the generated code's syntax and programming patterns. The impact of quantization on LLM-generated code logic is highly dependent on the model, problem size, and task the model is being applied to. We recommend testing LLMs on different quantization settings and on different code generation tasks before using it to write production code.}
  \end{minipage}
}

\subsection{Subjective Analysis of RQ1 and RQ2}

\update{Several findings from both Table \ref{tab:evals_plus} and \ref{tab:codebleu} warrant discussion. Results showed mixed effects from applying activation-aware quantization (AWQ) to our LLMs~\cite{AWQ_2023}. 
Normalizing the compression range to commonly activated layer weights may over-emphasize some weights.
This can boost performance on certain code generation tasks but hinder it on others. 
We have not empirically verified this, but it is a reasonable hypothesis given AWQ's mechanism.}

Certain code-specific LLMs outperform others due to differences in architecture, optimization, and training.
Mistral Instruct 7B likely benefits from sliding window pretraining and caching, enabling a larger input context. 
WizardCoder 7B's performance may stem from adversarial fine-tuning on progressively harder programming problems. 
StarCoder 2 7B likely suffers from lack of prompt-based fine-tuning, and CodeLlama 7B may require prompt engineering, which was not supported in this study.

\update{\subsection{\textbf{RQ3: Code quality issues}}
\label{sec:RQ3}

To evaluate the quality of LLM-generated code without prompt engineering, we conducted a static analysis using SonarQube \update{on the generated quantized and unquantized code} from RQ1 and RQ2. As seen in Table \ref{tab:sonarqube_results}, the SonarQube analysis covered a total of \update{6,756} Python files, revealing \update{1,848} detected issues from the LLM generated code. SonarQube's projection suggests that addressing all identified issues would require approximately \update{33 days} of development effort.

\begin{table}[h]
\centering
\scriptsize
\caption{\update{SonarQube Analysis Results}}
\begin{tabular}{|l|c|}
\hline
\textbf{Metric} & \textbf{Value} \\
\hline
Generated Python Solutions & 6,756 \\
Issues Detected & 1,848 \\
Estimated Remediation Cost & 33 days \\
\hline
\end{tabular}
\label{tab:sonarqube_results}
\end{table}

\subsubsection{Clean Code Attribute Analysis}

\begin{figure}[htbp]
\centering
\caption{Clean Code Attributes Issues}

\begin{tikzpicture}[scale=0.70]
\begin{axis}[
    ybar,
    bar width=0.8cm,
    symbolic x coords={Consistency, Intentionality, Adaptability, Responsibility},
    ylabel={Frequency of Issue},
    ymin=0,
    enlarge x limits=0.15,
    nodes near coords,
    nodes near coords align={vertical},
    xticklabel style={font=\footnotesize, rotate=45, anchor=north east},
    xtick style={draw=none},
    ]
\addplot coordinates {(Consistency, 577) (Intentionality, 1102) (Adaptability, 174) (Responsibility, 0)};
\end{axis}
\end{tikzpicture}
\label{fig:bar_chart_attributes}
\end{figure}

\update{Figure~\ref{fig:bar_chart_attributes} illustrates the distribution of 1,848 static analysis issues categorized by clean code attributes, as defined in Section~\ref{sec:sonarqube analysis}. Of the 1,848 issues, 1,102 (59.6\%) were related to \textit{intentionality}, 577 (31.2\%) to \textit{consistency}, 174 (9.4\%) to \textit{adaptability}, and none were associated with \textit{responsibility}.}
SonarQube estimates $\sim$213 hours of effort required to remedy the consistency issues, $\sim$288 hours for the intentionality issues, and $\sim$264 hours for adaptability issues.

\update{Comparing issues to total solutions, we observe that approximately \textbf{27.4\%} of the code samples contain at least one issue, indicating a substantial likelihood of encountering quality concerns in LLM generated outputs. The predominance of intentionality related issues suggests that LLMs frequently generate code that lacks clarity, logical structure, or completeness.}

\update{Consistency issues, while less prevalent, still account for nearly a third of all detected problems. These issues reflect a lack of uniform formatting, conventional naming, or recognizable structural patterns, traits that are essential for collaborative and maintainable software development. Inconsistencies like these in production environment can degrade readability and complicate integration, especially in multi-developer environments or long term projects.}

\update{Adaptability related issues were the least common among the identified categories, but their presence still signals potential barriers to maintainability. Code like this which is not modular, testable, or logically segmented can become brittle under evolving requirements, making it difficult to safely extend or reuse.}

\update{In the results we observe no violations related to responsibility. This absence is likely due to the nature of the coding tasks supplied to the models. Basic Python problems from our selected datasets should not invoke ethical or legal implications.}

\update{Overall, these results suggest that while LLMs demonstrate capabilities in code generation, significant limitations remain in their ability to produce clean, robust, and maintainable code by professional software engineering standards.}

\subsubsection{Software Quality Issues Analysis}

\begin{figure}[htbp]
\centering
\caption{Software Quality Issues}
\begin{tikzpicture}[scale=0.70]
\begin{axis}[
    ybar,
    bar width=0.8cm,
    ylabel={Frequency of Issue},
    ymin=0,
    enlarge x limits=0.3,
    xtick={1,2,3},
    xticklabels={Security, Reliability, Maintainability},
    nodes near coords,
    nodes near coords align={vertical},
    xtick style={draw=none},
    ]
\addplot coordinates {(1, 3) (2, 282) (3, 1574)};
\end{axis}
\end{tikzpicture}
\label{fig:bar_chart_software_quality}
\end{figure}

\update{Figure~\ref{fig:bar_chart_software_quality} presents the classification of the 1,848 identified issues from the generated Python code in terms of software quality, as defined in Section~\ref{sec:sonarqube analysis}.}

\update{The analysis reveals that out of the 1,848 issues, only 3 are security related, 282 are related to reliability, and the remaining 1,574 are categorized as maintainability concerns. This breakdown underscores a significant skew: approximately \textbf{85.2\%} of all reported issues fall under maintainability. That is, when LLMs generate problematic code, over three quarters of the time it reflects structural or stylistic deficiencies rather than outright defects or vulnerabilities. From a broader perspective, this means that \textbf{22.2\%} of all code samples produced by the tested LLMs exhibit at least one maintainability issue, i.e., a code smell.}
\update{SonarQube estimates $\sim$15 minutes of effort required to remedy the security issues, $\sim$98 hours for the reliability issues, and $\sim$672 hours for maintainability issues.
}

\update{The high prevalence of maintainability issues has important implications for software engineering. Code smells, while not immediately harmful, tend to accumulate technical debt \cite{cunningham_wycash_nodate} and reduce long-term productivity by making code harder to read, extend, and refactor. These issues often signal insufficient modularity, poor naming conventions, or redundant logic, areas where LLMs may mimic surface level syntax but lack deeper architectural understanding.}

\update{Overall, these results highlight that while these LLMs can generate working code, they often do so without adhering to best practices in maintainable software design.} 

\subsubsection{Generated Code Violations}

\begin{table}[h]
    \centering
    \caption{\update{Overview of SonarQube Rule Violations}}
    \scriptsize
    \begin{tabular}{|p{0.6\columnwidth}|c|}
    \toprule
    \textbf{SonarQube Rule} & \textbf{Number of Violations} \\
    \midrule
    Function names should comply with a naming convention. & 404 \\
    \hline
    Unused local variables should be removed. & 262 \\
    \hline
    Sections of code should not be commented out. & 225 \\
    \hline
    Floating-point numbers should not be tested for equality. & 170 \\
    \hline
    Built-ins should not be overshadowed by local variables. & 139 \\
    \hline
    String literals should not be duplicated. & 106 \\
    \hline
    \texttt{isinstance()} should be preferred over direct type comparisons. & 86 \\
    \hline
    Cognitive complexity of functions should not be too high. & 63 \\
    \hline
    The number and names of arguments passed to a function should match its parameters. & 49 \\
    \hline
    Unused function parameters should be removed. & 49 \\
    \hline
    The \texttt{print} statement should not be used. & 44 \\
    \hline
    Local variable and function parameter names should comply with a naming convention. & 41 \\
    \hline
    Mergeable \texttt{if} statements should be combined. & 37 \\
    \hline
    Uses of \texttt{TODO} tags should be tracked. & 34 \\
    \hline
    Conditional expressions should not be nested. & 22 \\
    \hline
    \texttt{SystemExit} should be re-raised. & 19 \\
    \hline
    All branches in a conditional structure should not have exactly the same implementation. & 15 \\
    \hline
    Non-empty statements should change control flow or have at least one side effect. & 15 \\
    \hline
    Variables should not be self-assigned. & 10 \\
    \hline
    Unused scope-limited definitions should be removed. & 7 \\
    \hline
    Calls should not be made to non-callable values. & 5 \\
    \hline
    Regular expression quantifiers and character classes should be used concisely. & 5 \\
    \bottomrule
    \end{tabular}
    \vspace{0.5em}
    
    {\footnotesize Note: Only rules with 5 or more violations are included in this table.}
    
    \label{tab:fix_recommendations}
\end{table}

\update{Table~\ref{tab:fix_recommendations} provides a detailed breakdown of the most frequently violated SonarQube rules found in the generated Python code, with only rules having five or more violations included. The data highlights several key trends in how LLMs handle code generation, particularly in regard to style, maintainability, and correctness.}

\update{The most prevalent issue, \textit{Function names should comply with a naming convention} (404 occurrences), indicates a systemic inconsistency in how function identifiers are produced. This suggests that while these LLMs can generate syntactically correct functions, they do not reliably adhere to common naming conventions such as snake\_case in Python, potentially hampering readability and cohesion across large codebases.}

\update{The second most common violation, \textit{Unused local variables should be removed} (262 occurrences), reflects inefficiencies in the generated code. These superfluous variables may result from incomplete or overly generic code patterns, and their presence can obscure program logic, making debugging and maintenance more difficult.}

\update{The third-ranked issue, \textit{Sections of code should not be commented out} (225 occurrences), stems from the LLMs including redundant or alternate code suggestions in comments. While possibly well intentioned (e.g., offering alternatives or placeholders), these patterns contribute to code clutter and reduce clarity.}

Several rules point to concerns in correctness and reliability:
\begin{itemize}[leftmargin=1em]
    \item \textit{Floating-point numbers should not be tested for equality} (170 occurrences) reveals a lack of numerical precision awareness, a subtle but critical flaw in scientific or financial computations.
    \item \textit{Built-ins should not be overshadowed by local variables} (139 occurrences) shows that LLMs occasionally redefine Python built-in names (e.g., \texttt{list}, \texttt{str}), potentially introducing unexpected runtime behavior.
    \item \textit{Cognitive complexity of functions should not be too high} (63 occurrences) suggests that LLMs generate overly intricate logic, which can hinder readability, testability, and refactorability.
\end{itemize}

Maintainability is also challenged by duplicated code patterns and poor abstraction:
    \textit{String literals should not be duplicated} (106 occurrences) and
    \textit{Conditional expressions should not be nested} (22 occurrences)
both signal that code structure could benefit from refactoring or decomposition into helper functions or constants.

The presence of 44 violations of the rule \textit{The \texttt{print} statement should not be used} and 34 uses of \textit{TODO} tags displays that LLMs frequently generate code with temporary debugging aids or placeholders practices discouraged in production environments.

The data from Table~\ref{tab:fix_recommendations} underscores a pattern of stylistic and structural code quality issues, rather than egregious security or logic flaws. These issues, while not catastrophic, accumulate as technical debt and can compromise maintainability in larger systems. 

\subsubsection{Manual Analysis}

In addition to automated static analysis using SonarQube, manual inspection was conducted on the code generated by each model across both benchmark datasets. This analysis provided further qualitative insights into recurring issues that were either missed or not fully captured by automated tooling.

\textbf{CodeLlama}:
In the HumanEvalPlus dataset, CodeLlama frequently generated repeated function definitions, often producing redundant logic with multiple identical or near-identical function implementations. Some responses included incomplete placeholder logic such as \texttt{\# TODO: Implement this function} followed by an empty \texttt{pass} statement, indicating a lack of meaningful attempt to solve the problem. In the MBPP Plus dataset, the most notable issue was the presence of infinite \texttt{print} loops, suggesting a failure in termination logic.

\textbf{Mistral Instruct}:
Mistral’s outputs in the HumanEvalPlus dataset exhibited frequent repetition of function definitions similar to CodeLlama. In the MBPP Plus dataset, the most prominent issue was the inclusion of commented out test cases or code blocks. While this may suggest a conservative code generation strategy, it can also lead to confusion and unusable output for downstream users.

\textbf{Starcoder2}:
Starcoder2 exhibited a mix of structural and syntactic issues in the HumanEvalPlus dataset, including invalid syntax such as broken import paths and improper file or directory references. It also frequently produced README templates or explanatory markdown content instead of runnable code. As with other models, repetition of function definitions was common. In the MBPP Plus dataset, Starcoder2 displayed similar problems, including invalid syntax and infinite \texttt{print} tests, resulting in non-terminating output. Repetitive code structures were also prevalent.

\textbf{WizardCoder}:
WizardCoder had the most diverse range of issues across datasets. In the HumanEvalPlus dataset, it often generated infinite \texttt{print} statements, and like other models, repeated function definitions. In the MBPP Plus dataset, WizardCoder exhibited several concerning behaviors, including hardcoded outputs that bypassed the actual problem logic, repeated function implementations, infinite \texttt{print} tests, and commented-out code blocks, all of which significantly reduced code utility and maintainability.

Across all models and datasets, a common theme was the repetition of function definitions, suggesting that these LLMs lack effective mechanisms for recognizing completion and avoiding redundancy. The inclusion of placeholders, infinite loops, commented-out code, and hardcoded values further illustrates the need for output validation in code generation systems. These manual findings complement the static analysis results, offering a more holistic picture of model performance and real-world applicability.

\subsubsection{Quantized vs. Unquantized Model Performance}

\begin{table}[h]
    \centering
    \caption{SonarQube Issues by Model}
    \scriptsize
    \begin{tabular}{p{0.8cm}p{0.5cm}p{0.2cm}p{0.5cm}p{0.2cm}p{0.5cm}p{0.2cm}p{0.5cm}p{0.2cm}p{0.4cm}}
    \toprule
     Model & \multicolumn{2}{p{1.3cm}}{CodeLlama 7B} & \multicolumn{2}{p{1.6cm}}{Mistral Instruct 7B} & \multicolumn{2}{p{1.4cm}}{WizardCoder 3B} & \multicolumn{2}{p{1.2cm}}{Starcoder2 7B} & \textbf{Total} \\
     \toprule
     Benchmark & Human EvalPlus & MBPP Plus & Human EvalPlus & MBPP Plus & Human EvalPlus & MBPP Plus & Human EvalPlus & MBPP Plus & \\
\midrule
4 bit        & 38 & 104 & 29 & 193 & 20 & 130 & 63 & 193 & \textbf{770} \\
8 bit        & 20 & 70  & 48 & 176 & 36 & 98  & 29 & 30  & \textbf{507} \\
Unquantized  & 28 & 137 & 23 & 107 & 49 & 105 & 48 & 74  & \textbf{571} \\
\midrule
\textbf{Total} & \textbf{86} & \textbf{311} & \textbf{100} & \textbf{476} & \textbf{105} & \textbf{333} & \textbf{140} & \textbf{297} & \textbf{1,848} \\
\bottomrule
\end{tabular}
    \label{tab:sonar_model_results}
\end{table}

Table~\ref{tab:sonar_model_results} provides a comparative analysis of SonarQube reported code quality issues across the four different LLMs included in this study at three quantization levels. The results are aggregated over two benchmark datasets: HumanEvalPlus and MBPP Plus.

From the table, it is clear that 4-bit quantization significantly impacts the quality of code generated by LLMs, with 4-bit quantized models producing the most issues overall (770 issues), followed by unquantized models (571 issues), and 8-bit quantized models producing the fewest issues (507). This contradicts the commonly held assumption that unquantized models consistently yield higher-quality outputs. Instead, 8-bit quantization appears to strike a balance between model compression and code generation quality.

A look at the per-model breakdown reveals further insights:
\begin{itemize}[leftmargin=1em]
    \item \textbf{CodeLlama 7B} shows a noticeable improvement when moving from 4-bit (142 issues) to 8-bit (90 issues), with a slight increase in issues when unquantized (165 issues), particularly in the MBPP Plus dataset. This suggests that while compression helps reduce certain redundancies, unquantized models may overgenerate or introduce more nuanced issues.
    \item \textbf{Mistral Instruct 7B} performs best in its unquantized form on HumanEvalPlus (23 issues), but worst in 4-bit MBPP Plus (193 issues), showing strong sensitivity to quantization on that dataset.  
    \item \textbf{WizardCoder 7B} has a surprisingly high number of issues in its unquantized form on HumanEvalPlus (49 issues), indicating that smaller models, even when fully unquantized, may struggle with code precision. Interestingly, the 4-bit version performs better than unquantized in this case (20 issues).
    \item \textbf{Starcoder2 7B} stands out for its high issue count in the 4-bit and unquantized MBPP Plus datasets (193 and 74 issues, respectively), compared to only 30 issues in the 8-bit version. This suggests that 8-bit quantization may help this model avoid certain structural issues or overfitting behaviors.
\end{itemize}

When considering total issues by dataset, MBPP Plus contributes the majority (1,417 out of 1,848 issues), reinforcing that it is more sensitive to LLM generation flaws, potentially due to its broader coverage of algorithmic challenges and formatting expectations compared to HumanEvalPlus (431 issues).

In summary, Table~\ref{tab:sonar_model_results} demonstrates that while 4-bit quantization can substantially degrade code quality, 8-bit quantization often preserves or even enhances performance relative to unquantized baselines. These findings have practical implications for deploying LLMs in resource-constrained environments without compromising output quality, particularly in code-generation use cases.

\colorbox{gray!20}{%
  \begin{minipage}{0.93\columnwidth}
    \textbf{RQ3 Findings:} 
    We evaluated LLM-generated code using SonarQube and manual inspection. SonarQube identified 1,848 issues ($\approx$33 days remediation), with 85.2\% under \textit{maintainability}, reflecting stylistic and structural flaws rather than security or reliability risks. Clean code analysis showed most issues stemmed from \textit{intentionality} (59.6\%), followed by \textit{consistency} (31.2\%) and \textit{adaptability} (9.4\%), with none tied to \textit{responsibility}. Violations included inconsistent naming, unused variables, commented-out code, and misuse of built-ins, factors that add technical debt and hinder readability. Manual review revealed duplicated functions, incomplete logic, and infinite loops. Overall, smaller LLMs produce functional code but often fall short of professional standards for clarity and maintainability.
  \end{minipage}
}}

\section{Threats to Validity}

This section outlines potential threats to the validity of our findings. 
\indent \textbf{\textit{Internal Validity.}}
Our results may be influenced by quantization implementation details, as we used Activation Aware Weight Quantization (AWQ). Misconfigurations could skew performance, and testing all models on the same hardware may bias results toward models optimized for that hardware.
Another threat is the stochastic nature of LLM outputs. Even with consistent settings, generation variability can affect benchmark outcomes, particularly where small code differences impact passing rates.

\textbf{\textit{External Validity.}}
Findings may not generalize beyond the smaller open-source LLMs tested. Larger  proprietary models like GPT-5, or those trained on different datasets, may behave differently. Our reliance on two Python-focused  benchmarks (HumanEvalPlus and MBPP Plus) also limits generalizability to other programming languages or real-world tasks. A further risk is data leakage, as benchmark problems may overlap with LLM training data. While we used standard evaluation datasets, the lack of transparency in training corpora prevents full verification.

\textbf{\textit{Construct Validity.}}
Evaluation relied on benchmark pass rates and CodeBLEU. These standard metrics capture correctness and similarity but do not fully reflect maintainability, readability, or scalability. CodeBLEU, in particular, focuses on syntactic similarity rather than functional quality.
Similarly, SonarQube provided useful insights into code smells and maintainability, but static analysis can not capture all runtime or functional issues.

\textbf{\textit{Conclusion Validity.}}
Our conclusion is limited by the number of models (four) and quantization settings (three) tested. Results may differ on other hardware architectures. 
In addition, we evaluated only one generated solution per prompt, whereas sampling multiple outputs or using feedback could improve performance, suggesting our results may underestimate real-world potential.
\section{Conclusion}

\update{This study highlights the dual promises and limitations of smaller LLMs in code generation as a proof of concept. These models can mimic human-like syntax and implement basic logic without prompt engineering but struggle with larger or more complex problems. 
Quantization showed mixed effects, improving test passing rates in some models while lowering them on others.}

\update{Additionally, LLM generated code quality can suffer from consistency and maintainability issues, even if it works properly. These include not aligning with recommended code or syntax practices, code smells that lead to technical debt, potential security issues, and lack of proper comments. Interestingly, quantization tends to reduce these issues for some LLMs and significantly in some particular quantization settings and applications. The varied effects of quantization should be studied further, both in research and industry uses, to investigate potential causes and remedies.}

\update{Future work should focus on further ways to evaluate different code-specific LLMs for the quality, functionality, and usefulness of generated code. Focusing further on the effects of quantization would also be helpful for research purposes. Establishing explanations for why certain LLMs perform better or worse in specific scenarios during quantization, as well as ways for mitigating negative effects and emphasizing positive ones, should be addressed. Further work should also quantify the effects of prompt engineering on LLM-generated code quality, as well as various techniques to improve that code quality in both functionality and compliance with recommended standards.}

\bibliographystyle{ACM-Reference-Format}
\bibliography{references}

@online{sonarqube-doc,
  title = {SonarQube Documentation},
  author = {{SonarSource}},
  year = {latest},
  url = {https://docs.sonarsource.com/sonarqube/latest/user-guide/clean-code/definition/},
}

@article{chen2021evaluating,
  title={Evaluating large language models trained on code},
  author={Chen, Mark and Tworek, Jerry and Jun, Heewoo and Yuan, Qiming and Pinto, Henrique Ponde De Oliveira and Kaplan, Jared and Edwards, Harri and Burda, Yuri and Joseph, Nicholas and Brockman, Greg and others},
  journal={arXiv preprint arXiv:2107.03374},
  year={2021}
}

@article{siddiq2023generate,
  title={Generate and pray: Using sallms to evaluate the security of llm generated code},
  author={Siddiq, Mohammed Latif and Santos, Joanna CS},
  journal={arXiv preprint arXiv:2311.00889},
  year={2023}
}

@article{Siddiq_Santos_2023, title={Generate and Pray: Using SALLMS to Evaluate the Security of LLM Generated Code}, url={http://arxiv.org/abs/2311.00889}, abstractNote={With the growing popularity of Large Language Models (e.g., GitHub Copilot, ChatGPT, etc.) in software engineers’ daily practices, it is important to ensure that the code generated by these tools is not only functionally correct but also free of vulnerabilities. Although LLMs can help developers to be more productive, prior empirical studies have shown that LLMs can generate insecure code. There are two contributing factors to the insecure code generation. First, existing datasets used to evaluate Large Language Models (LLMs) do not adequately represent genuine software engineering tasks sensitive to security. Instead, they are often based on competitive programming challenges or classroom-type coding tasks. In real-world applications, the code produced is integrated into larger codebases, introducing potential security risks. There’s a clear absence of benchmarks that focus on evaluating the security of the generated code. Second, existing evaluation metrics primarily focus on the functional correctness of the generated code while ignoring security considerations. Metrics such as pass@k gauge the probability of obtaining the correct code in the top k suggestions. Other popular metrics like BLEU, CodeBLEU, ROUGE, and METEOR similarly emphasize functional accuracy, neglecting security implications. In light of these research gaps, in this paper, we described SALLM, a framework to benchmark LLMs’ abilities to generate secure code systematically. This framework has three major components: a novel dataset of security-centric Python prompts, an evaluation environment to test the generated code, and novel metrics to evaluate the models’ performance from the perspective of secure code generation.}, note={arXiv:2311.00889 [cs]}, number={arXiv:2311.00889}, publisher={arXiv}, author={Siddiq, Mohammed Latif and Santos, Joanna C. S.}, year={2023}, month=nov, language={en} }

@article{Du_Luu_Ji_Ng_2024, title={Mercury: An Efficiency Benchmark for LLM Code Synthesis}, url={http://arxiv.org/abs/2402.07844}, DOI={10.48550/arXiv.2402.07844}, abstractNote={Despite advancements in evaluating Large Language Models (LLMs) for code synthesis, benchmarks have predominantly focused on functional correctness, overlooking the importance of code efficiency. We present Mercury, the first benchmark designated for assessing the code efficiency of LLM code synthesis tasks. Mercury consists of 1,889 programming tasks covering diverse difficulty levels alongside test case generators generating unlimited cases for comprehensive evaluation. Unlike existing benchmarks, Mercury integrates a novel metric Beyond@K to measure normalized code efficiency based on historical submissions, leading to a new evaluation indicator for code synthesis, which encourages generating functionally correct and computationally efficient code, mirroring the real-world software development standard. Our findings reveal that while LLMs demonstrate the remarkable capability to generate functionally correct code, there still exists a substantial gap in their efficiency output, underscoring a new frontier for LLM research and development.}, note={arXiv:2402.07844 [cs]}, number={arXiv:2402.07844}, publisher={arXiv}, author={Du, Mingzhe and Luu, Anh Tuan and Ji, Bin and Ng, See-Kiong}, year={2024}, month=feb }

@article{Bhatt_Chennabasappa_2023, title={Purple Llama CyberSecEval: A Secure Coding Benchmark for Language Models}, url={http://arxiv.org/abs/2312.04724}, DOI={10.48550/arXiv.2312.04724}, abstractNote={This paper presents CyberSecEval, a comprehensive benchmark developed to help bolster the cybersecurity of Large Language Models (LLMs) employed as coding assistants. As what we believe to be the most extensive unified cybersecurity safety benchmark to date, CyberSecEval provides a thorough evaluation of LLMs in two crucial security domains: their propensity to generate insecure code and their level of compliance when asked to assist in cyberattacks. Through a case study involving seven models from the Llama 2, Code Llama, and OpenAI GPT large language model families, CyberSecEval effectively pinpointed key cybersecurity risks. More importantly, it offered practical insights for refining these models. A significant observation from the study was the tendency of more advanced models to suggest insecure code, highlighting the critical need for integrating security considerations in the development of sophisticated LLMs. CyberSecEval, with its automated test case generation and evaluation pipeline covers a broad scope and equips LLM designers and researchers with a tool to broadly measure and enhance the cybersecurity safety properties of LLMs, contributing to the development of more secure AI systems.}, note={arXiv:2312.04724 [cs]}, number={arXiv:2312.04724}, publisher={arXiv}, author={Bhatt, Manish and Chennabasappa, Sahana and Nikolaidis, Cyrus and Wan, Shengye and Evtimov, Ivan and Gabi, Dominik and Song, Daniel and Ahmad, Faizan and Aschermann, Cornelius and Fontana, Lorenzo and Frolov, Sasha and Giri, Ravi Prakash and Kapil, Dhaval and Kozyrakis, Yiannis and LeBlanc, David and Milazzo, James and Straumann, Aleksandar and Synnaeve, Gabriel and Vontimitta, Varun and Whitman, Spencer and Saxe, Joshua}, year={2023}, month=dec }

@article{Zhong_Wang_2024, title={Can ChatGPT replace StackOverflow? A Study on Robustness and Reliability of Large Language Model Code Generation}, url={http://arxiv.org/abs/2308.10335}, DOI={10.48550/arXiv.2308.10335}, abstractNote={Recently, the large language models (LLMs) have shown extraordinary ability in understanding natural language and generating programming code. It has been a common practice of software engineers to consult LLMs when encountering coding questions. Although efforts have been made to avoid syntax errors and align the code with the intended semantics, the reliability and robustness of the code generationfrom LLMs have not yet been thoroughly studied. The executable code is not equivalent to the reliable and robust code, especially in the context of real-world software development. The misuse of APIs in the generated code could lead to severe problem, such as resource leaks, program crashes. To make things worse, the users of LLM code generation services are actually the developers that are most vulnerable to these code that seems right -- They are always novice developers that are not familiar with the APIs that LLMs generate code for them. Therefore, they could hardly tell the misuse in the code generated by LLMs, which further facilitates the incorrect code applied in real-world software. Existing code evaluation benchmark and datasets focus on crafting small tasks such as programming questions in coding interviews, which however deviates from the problem that developers would ask LLM for real-world coding help. To fill the missing piece, in this work, we propose a dataset RobustAPI for evaluating the reliability and robustness of code generated by LLMs. We collect 1208 coding questions from StackOverflow on 24 representative Java APIs. We summarize thecommon misuse patterns of these APIs and evaluate them oncurrent popular LLMs. The evaluation results show that evenfor GPT-4, 62% of the generated code contains API misuses,which would cause unexpected consequences if the code isintroduced into real-world software.}, note={arXiv:2308.10335 [cs]}, number={arXiv:2308.10335}, publisher={arXiv}, author={Zhong, Li and Wang, Zilong}, year={2024}, month=jan }

@inproceedings{Vaithilingam_2022,
author = {Vaithilingam, Priyan and Zhang, Tianyi and Glassman, Elena L.},
title = {Expectation vs. Experience: Evaluating the Usability of Code Generation Tools Powered by Large Language Models},
year = {2022},
isbn = {9781450391566},
publisher = {Association for Computing Machinery},
address = {New York, NY, USA},
url = {https://doi.org/10.1145/3491101.3519665},
doi = {10.1145/3491101.3519665},
booktitle = {Extended Abstracts of the 2022 CHI Conference on Human Factors in Computing Systems},
articleno = {332},
numpages = {7},
location = {New Orleans, LA, USA},
series = {CHI EA '22}
}

@article{Nguyen_Babe_Zi_Guha_Anderson_Feldman_2024, title={How Beginning Programmers and Code LLMs (Mis)read Each Other}, url={http://arxiv.org/abs/2401.15232}, DOI={10.48550/arXiv.2401.15232}, abstractNote={Generative AI models, specifically large language models (LLMs), have made strides towards the long-standing goal of text-to-code generation. This progress has invited numerous studies of user interaction. However, less is known about the struggles and strategies of non-experts, for whom each step of the text-to-code problem presents challenges: describing their intent in natural language, evaluating the correctness of generated code, and editing prompts when the generated code is incorrect. This paper presents a large-scale controlled study of how 120 beginning coders across three academic institutions approach writing and editing prompts. A novel experimental design allows us to target specific steps in the text-to-code process and reveals that beginners struggle with writing and editing prompts, even for problems at their skill level and when correctness is automatically determined. Our mixed-methods evaluation provides insight into student processes and perceptions with key implications for non-expert Code LLM use within and outside of education.}, note={arXiv:2401.15232 [cs]}, number={arXiv:2401.15232}, publisher={arXiv}, author={Nguyen, Sydney and Babe, Hannah McLean and Zi, Yangtian and Guha, Arjun and Anderson, Carolyn Jane and Feldman, Molly Q.}, year={2024}, month=jan }

@misc{ren2020codebleu,
      title={CodeBLEU: a Method for Automatic Evaluation of Code Synthesis}, 
      author={Shuo Ren and Daya Guo and Shuai Lu and Long Zhou and Shujie Liu and Duyu Tang and Neel Sundaresan and Ming Zhou and Ambrosio Blanco and Shuai Ma},
      year={2020},
      eprint={2009.10297},
      archivePrefix={arXiv},
      primaryClass={cs.SE}
}

@INPROCEEDINGS{Bender2020-ck,
  title     = "Climbing towards {NLU}: On Meaning, Form, and Understanding in
               the Age of Data",
  author    = "Bender, Emily M and Koller, Alexander",
  booktitle = "Association for Computational Linguistics",
  pages     = "5185--5198",
  year      =  2020
}

@inproceedings{de2024automatic,
  title={Automatic Assessment of Architectural Anti-patterns and Code Smells in Student Software Projects},
  author={De Luca, Marco and Di Meglio, Sergio and Fasolino, Anna Rita and Starace, Luigi Libero Lucio and Tramontana, Porfirio},
  booktitle={Proceedings of the 28th International Conference on Evaluation and Assessment in Software Engineering},
  pages={565--569},
  year={2024}
}

@article{cunningham_wycash_nodate,
  title={The WyCash portfolio management system},
  author={Cunningham, Ward},
  journal={ACM Sigplan Oops Messenger},
  volume={4},
  number={2},
  pages={29--30},
  year={1992},
  publisher={ACM New York, NY, USA}
}

@inproceedings{gurung2024static,
  title={Static code analysis for reducing energy code smells in different loop types: a case study in Java},
  author={Gurung, Ram Prasad and Porras, Jari and Koistinaho, Jarkko},
  booktitle={2024 10th International Conference on ICT for Sustainability (ICT4S)},
  pages={292--302},
  year={2024},
  organization={IEEE}
}

@inproceedings{mat2024enhancing,
  title={Enhancing Code Quality Through Static Analysis: Optimizing the Flava Tool for Detecting Code Smells and Errors Using SonarQube Integration},
  author={Mat'a{\v{s}}ov{\'a}, Sylvia and Chovanec, Martin and Rusn{\'a}kov{\'a}, Ren{\'a}ta and {\v{C}}atloch, Du{\v{s}}an},
  booktitle={2024 International Conference on Emerging eLearning Technologies and Applications (ICETA)},
  pages={431--438},
  year={2024},
  organization={IEEE}
}

@article{Luo_Xu_Zhao_Sun_Geng_Hu_Tao_Ma_Lin_Jiang_2023, title={WizardCoder: Empowering Code Large Language Models with Evol-Instruct}, url={http://arxiv.org/abs/2306.08568}, abstractNote={Code Large Language Models (Code LLMs), such as StarCoder, have demonstrated exceptional performance in code-related tasks. However, most existing models are solely pre-trained on extensive raw code data without instruction finetuning. In this paper, we introduce WizardCoder, which empowers Code LLMs with complex instruction fine-tuning, by adapting the Evol-Instruct method to the domain of code. Through comprehensive experiments on four prominent code generation benchmarks, namely HumanEval, HumanEval+, MBPP, and DS1000, we unveil the exceptional capabilities of our model. It surpasses all other open-source Code LLMs by a substantial margin. Moreover, our model even outperforms the largest closed LLMs, Anthropic’s Claude and Google’s Bard, on HumanEval and HumanEval+. Our code, model weights, and data are public at https://github.com/nlpxucan/WizardLM.}, note={arXiv:2306.08568 [cs]}, number={arXiv:2306.08568}, publisher={arXiv}, author={Luo, Ziyang and Xu, Can and Zhao, Pu and Sun, Qingfeng and Geng, Xiubo and Hu, Wenxiang and Tao, Chongyang and Ma, Jing and Lin, Qingwei and Jiang, Daxin}, year={2023}, month=jun, language={en} }

@article{Mistral_2023, title={Mistral 7B}, url={http://arxiv.org/abs/2310.06825}, abstractNote={We introduce Mistral 7B v0.1, a 7-billion-parameter language model engineered for superior performance and efficiency. Mistral 7B outperforms Llama 2 13B across all evaluated benchmarks, and Llama 1 34B in reasoning, mathematics, and code generation. Our model leverages grouped-query attention (GQA) for faster inference, coupled with sliding window attention (SWA) to effectively handle sequences of arbitrary length with a reduced inference cost. We also provide a model fine-tuned to follow instructions, Mistral 7B -- Instruct, that surpasses the Llama 2 13B -- Chat model both on human and automated benchmarks. Our models are released under the Apache 2.0 license.}, note={arXiv:2310.06825 [cs]}, number={arXiv:2310.06825}, publisher={arXiv}, author={Jiang, Albert Q. and Sablayrolles, Alexandre and Mensch, Arthur and Bamford, Chris and Chaplot, Devendra Singh and Casas, Diego de las and Bressand, Florian and Lengyel, Gianna and Lample, Guillaume and Saulnier, Lucile and Lavaud, Lélio Renard and Lachaux, Marie-Anne and Stock, Pierre and Scao, Teven Le and Lavril, Thibaut and Wang, Thomas and Lacroix, Timothée and Sayed, William El}, year={2023}, month=oct, language={en} }

@article{StarCoder2_2024, title={StarCoder 2 and The Stack v2: The Next Generation}, url={http://arxiv.org/abs/2402.19173}, abstractNote={The BigCode project,1 an open-scientific collaboration focused on the responsible development of Large Language Models for Code (Code LLMs), introduces StarCoder2. In partnership with Software Heritage (SWH),2 we build The Stack v2 on top of the digital commons of their source code archive. Alongside the SWH repositories spanning 619 programming languages, we carefully select other high-quality data sources, such as GitHub pull requests, Kaggle notebooks, and code documentation. This results in a training set that is 4× larger than the first StarCoder dataset. We train StarCoder2 models with 3B, 7B, and 15B parameters on 3.3 to 4.3 trillion tokens and thoroughly evaluate them on a comprehensive set of Code LLM benchmarks.}, note={arXiv:2402.19173 [cs]}, number={arXiv:2402.19173}, publisher={arXiv}, author={Lozhkov, Anton and Li, Raymond and Allal, Loubna Ben and Cassano, Federico and Lamy-Poirier, Joel and Tazi, Nouamane and Tang, Ao and Pykhtar, Dmytro and Liu, Jiawei and Wei, Yuxiang and Liu, Tianyang and Tian, Max and Kocetkov, Denis and Zucker, Arthur and Belkada, Younes and Wang, Zijian and Liu, Qian and Abulkhanov, Dmitry and Paul, Indraneil and Li, Zhuang and Li, Wen-Ding and Risdal, Megan and Li, Jia and Zhu, Jian and Zhuo, Terry Yue and Zheltonozhskii, Evgenii and Dade, Nii Osae Osae and Yu, Wenhao and Krauß, Lucas and Jain, Naman and Su, Yixuan and He, Xuanli and Dey, Manan and Abati, Edoardo and Chai, Yekun and Muennighoff, Niklas and Tang, Xiangru and Oblokulov, Muhtasham and Akiki, Christopher and Marone, Marc and Mou, Chenghao and Mishra, Mayank and Gu, Alex and Hui, Binyuan and Dao, Tri and Zebaze, Armel and Dehaene, Olivier and Patry, Nicolas and Xu, Canwen and McAuley, Julian and Hu, Han and Scholak, Torsten and Paquet, Sebastien and Robinson, Jennifer and Anderson, Carolyn Jane and Chapados, Nicolas and Patwary, Mostofa and Tajbakhsh, Nima and Jernite, Yacine and Ferrandis, Carlos Muñoz and Zhang, Lingming and Hughes, Sean and Wolf, Thomas and Guha, Arjun and von Werra, Leandro and de Vries, Harm}, year={2024}, month=feb, language={en} }

@article{CodeLlama_2024, title={Code Llama: Open Foundation Models for Code}, url={http://arxiv.org/abs/2308.12950}, abstractNote={We release Code Llama, a family of large language models for code based on Llama 2 providing state-of-the-art performance among open models, infilling capabilities, support for large input contexts, and zero-shot instruction following ability for programming tasks. We provide multiple flavors to cover a wide range of applications: foundation models (Code Llama), Python specializations (Code Llama - Python), and instruction-following models (Code Llama - Instruct) with 7B, 13B, 34B and 70B parameters each. All models are trained on sequences of 16k tokens and show improvements on inputs with up to 100k tokens. 7B, 13B and 70B Code Llama and Code Llama - Instruct variants support infilling based on surrounding content. Code Llama reaches state-of-the-art performance among open models on several code benchmarks, with scores of up to 67% and 65% on HumanEval and MBPP, respectively. Notably, Code Llama - Python 7B outperforms Llama 2 70B on HumanEval and MBPP, and all our models outperform every other publicly available model on MultiPL-E. We release Code Llama under a permissive license that allows for both research and commercial use.}, note={arXiv:2308.12950 [cs]}, number={arXiv:2308.12950}, publisher={arXiv}, author={Rozière, Baptiste and Gehring, Jonas and Gloeckle, Fabian and Sootla, Sten and Gat, Itai and Tan, Xiaoqing Ellen and Adi, Yossi and Liu, Jingyu and Sauvestre, Romain and Remez, Tal and Rapin, Jérémy and Kozhevnikov, Artyom and Evtimov, Ivan and Bitton, Joanna and Bhatt, Manish and Ferrer, Cristian Canton and Grattafiori, Aaron and Xiong, Wenhan and Défossez, Alexandre and Copet, Jade and Azhar, Faisal and Touvron, Hugo and Martin, Louis and Usunier, Nicolas and Scialom, Thomas and Synnaeve, Gabriel}, year={2024}, month=jan, language={en} }

@article{AWQ_2023, title={AWQ: Activation-aware Weight Quantization for LLM Compression and Acceleration}, url={http://arxiv.org/abs/2306.00978}, abstractNote={Large language models (LLMs) have shown excellent performance on various tasks, but the astronomical model size raises the hardware barrier for serving (memory size) and slows down token generation (memory bandwidth). In this paper, we propose Activation-aware Weight Quantization (AWQ), a hardware-friendly approach for LLM low-bit weight-only quantization. Our method is based on the observation that weights are not equally important: protecting only 1% of salient weights can greatly reduce quantization error. We then propose to search for the optimal perchannel scaling that protects the salient weights by observing the activation, not weights. AWQ does not rely on any backpropagation or reconstruction, so it can well preserve LLMs’ generalization ability on different domains and modalities, without overfitting to the calibration set. AWQ outperforms existing work on various language modeling and domain-specific benchmarks. Thanks to better generalization, it achieves excellent quantization performance for instruction-tuned LMs and, for the first time, multi-modal LMs. Alongside AWQ, we implement an efficient and flexible inference framework tailored for LLMs on the edge, offering more than 3× speedup over the Huggingface FP16 implementation on both desktop and mobile GPUs. It also democratizes the deployment of the 70B Llama-2 model on mobile GPU (NVIDIA Jetson Orin 64GB).}, note={arXiv:2306.00978 [cs]}, number={arXiv:2306.00978}, publisher={arXiv}, author={Lin, Ji and Tang, Jiaming and Tang, Haotian and Yang, Shang and Dang, Xingyu and Gan, Chuang and Han, Song}, year={2023}, month=oct, language={en} }

@article{liu2024your,
  title={Is your code generated by chatgpt really correct? rigorous evaluation of large language models for code generation},
  author={Liu, Jiawei and Xia, Chunqiu Steven and Wang, Yuyao and Zhang, Lingming},
  journal={Advances in Neural Information Processing Systems},
  volume={36},
  year={2024}
}

@article{MBPP_2021, title={Program Synthesis with Large Language Models}, url={http://arxiv.org/abs/2108.07732}, abstractNote={This paper explores the limits of the current generation of large language models for program synthesis in general purpose programming languages. We evaluate a collection of such models (with between 244M and 137B parameters) on two new benchmarks, MBPP and MathQA-Python, in both the few-shot and ﬁne-tuning regimes. Our benchmarks are designed to measure the ability of these models to synthesize short Python programs from natural language descriptions. The Mostly Basic Programming Problems (MBPP) dataset contains 974 programming tasks, designed to be solvable by entry-level programmers. The MathQA-Python dataset, a Python version of the MathQA benchmark, contains 23914 problems that evaluate the ability of the models to synthesize code from more complex text. On both datasets, we ﬁnd that synthesis performance scales log-linearly with model size. Our largest models, even without ﬁnetuning on a code dataset, can synthesize solutions to 59.6% of the problems from MBPP using few-shot learning with a well-designed prompt. Fine-tuning on a held-out portion of the dataset improves performance by about 10 percentage points across most model sizes. On the MathQA-Python dataset, the largest ﬁne-tuned model achieves 83.8% accuracy. Going further, we study the model’s ability to engage in dialog about code, incorporating human feedback to improve its solutions. We ﬁnd that natural language feedback from a human halves the error rate compared to the model’s initial prediction. Additionally, we conduct an error analysis to shed light on where these models fall short and what types of programs are most difﬁcult to generate. Finally, we explore the semantic grounding of these models by ﬁne-tuning them to predict the results of program execution. We ﬁnd that even our best models are generally unable to predict the output of a program given a speciﬁc input.}, note={arXiv:2108.07732 [cs]}, number={arXiv:2108.07732}, publisher={arXiv}, author={Austin, Jacob and Odena, Augustus and Nye, Maxwell and Bosma, Maarten and Michalewski, Henryk and Dohan, David and Jiang, Ellen and Cai, Carrie and Terry, Michael and Le, Quoc and Sutton, Charles}, year={2021}, month=aug, language={en} }

@article{GPTQ_2023, title={GPTQ: Accurate Post-Training Quantization for Generative Pre-trained Transformers}, url={http://arxiv.org/abs/2210.17323}, abstractNote={Generative Pre-trained Transformer models, known as GPT or OPT, set themselves apart through breakthrough performance across complex language modelling tasks, but also by their extremely high computational and storage costs. Speciﬁcally, due to their massive size, even inference for large, highly-accurate GPT models may require multiple performant GPUs, which limits the usability of such models. While there is emerging work on relieving this pressure via model compression, the applicability and performance of existing compression techniques is limited by the scale and complexity of GPT models. In this paper, we address this challenge, and propose GPTQ, a new one-shot weight quantization method based on approximate second-order information, that is both highlyaccurate and highly-efﬁcient. Speciﬁcally, GPTQ can quantize GPT models with 175 billion parameters in approximately four GPU hours, reducing the bitwidth down to 3 or 4 bits per weight, with negligible accuracy degradation relative to the uncompressed baseline. Our method more than doubles the compression gains relative to previously-proposed one-shot quantization methods, preserving accuracy, allowing us for the ﬁrst time to execute an 175 billion-parameter model inside a single GPU for generative inference. Moreover, we also show that our method can still provide reasonable accuracy in the extreme quantization regime, in which weights are quantized to 2-bit or even ternary quantization levels. We show experimentally that these improvements can be leveraged for end-to-end inference speedups over FP16, of around 3.25x when using high-end GPUs (NVIDIA A100) and 4.5x when using more cost-effective ones (NVIDIA A6000). The implementation is available at https://github.com/IST-DASLab/gptq.}, note={arXiv:2210.17323 [cs]}, number={arXiv:2210.17323}, publisher={arXiv}, author={Frantar, Elias and Ashkboos, Saleh and Hoefler, Torsten and Alistarh, Dan}, year={2023}, month=mar, language={en} }

@article{LLM_SoftwareEngineeringSurvey_2023, title={Large Language Models for Software Engineering: Survey and Open Problems}, url={http://arxiv.org/abs/2310.03533}, abstractNote={This paper provides a survey of the emerging area of Large Language Models (LLMs) for Software Engineering (SE). It also sets out open research challenges for the application of LLMs to technical problems faced by software engineers. LLMs’ emergent properties bring novelty and creativity with applications right across the spectrum of Software Engineering activities including coding, design, requirements, repair, refactoring, performance improvement, documentation and analytics. However, these very same emergent properties also pose significant technical challenges; we need techniques that can reliably weed out incorrect solutions, such as hallucinations. Our survey reveals the pivotal role that hybrid techniques (traditional SE plus LLMs) have to play in the development and deployment of reliable, efficient and effective LLM-based SE.}, note={arXiv:2310.03533 [cs]}, number={arXiv:2310.03533}, publisher={arXiv}, author={Fan, Angela and Gokkaya, Beliz and Harman, Mark and Lyubarskiy, Mitya and Sengupta, Shubho and Yoo, Shin and Zhang, Jie M.}, year={2023}, month=nov, language={en} }

@article{SWE_Bench_2024, title={SWE-bench: Can Language Models Resolve Real-World GitHub Issues?}, url={http://arxiv.org/abs/2310.06770}, abstractNote={Language models have outpaced our ability to evaluate them effectively, but for their future development it is essential to study the frontier of their capabilities. We find real-world software engineering to be a rich, sustainable, and challenging testbed for evaluating the next generation of language models. To this end, we introduce SWE-bench, an evaluation framework consisting of 2,294 software engineering problems drawn from real GitHub issues and corresponding pull requests across 12 popular Python repositories. Given a codebase along with a description of an issue to be resolved, a language model is tasked with editing the codebase to address the issue. Resolving issues in SWE-bench frequently requires understanding and coordinating changes across multiple functions, classes, and even files simultaneously, calling for models to interact with execution environments, process extremely long contexts and perform complex reasoning that goes far beyond traditional code generation tasks. Our evaluations show that both state-ofthe-art proprietary models and our fine-tuned model SWE-Llama can resolve only the simplest issues. The best-performing model, Claude 2, is able to solve a mere 1.96% of the issues. Advances on SWE-bench represent steps towards LMs that are more practical, intelligent, and autonomous.}, note={arXiv:2310.06770 [cs]}, number={arXiv:2310.06770}, publisher={arXiv}, author={Jimenez, Carlos E. and Yang, John and Wettig, Alexander and Yao, Shunyu and Pei, Kexin and Press, Ofir and Narasimhan, Karthik}, year={2024}, month=apr, language={en} }

\end{document}